\documentclass[11pt,a4paper]{article}
\usepackage[hyperref]{emnlp2018}
\usepackage{times}
\usepackage{latexsym}
\usepackage{url}
\usepackage{graphicx}
\usepackage{subcaption}
\usepackage{amsmath, amssymb}
\usepackage[ruled,vlined]{algorithm2e}
\usepackage{algorithmic}

\aclfinalcopy 

\title{LRMM: Learning to Recommend with Missing Modalities}

\author{Cheng Wang \\
  NEC Laboratories Europe \\
  {\tt ~~~~~~~~~~~~~~~~~~~~~\{cheng.wang, mathias.niepert\}@neclab.eu} \\\And
  Mathias Niepert \\
  NEC Laboratories Europe \\
  \\ \And
  Hui Li\thanks{\;\;Work done in NEC Laboratories Europe.} \\
  The University of Hong Kong \\
  {\tt hli2@cs.hku.hk} \\}

\date{}

\begin{document}
\maketitle
\begin{abstract}
Multimodal learning has shown promising performance in content-based recommendation due to the auxiliary user and item information of multiple modalities such as text and images.  However, the problem of incomplete and missing modality is rarely explored and most existing methods fail in learning a recommendation model with missing or corrupted modalities. In this paper, we propose LRMM, a novel framework that mitigates not only the problem of missing modalities but also more generally the cold-start problem of recommender systems. We propose modality dropout (\textit{m}-drop) and a multimodal sequential autoencoder (\textit{m}-auto) to learn multimodal representations for complementing and imputing missing modalities. 
Extensive experiments on real-world Amazon data show that LRMM achieves state-of-the-art performance on rating prediction tasks. More importantly, LRMM is more robust to previous methods in alleviating data-sparsity and the cold-start problem.
\end{abstract}

\section{Introduction}

Recommender systems (RS) are useful filtering tools which aid customers in a personalized way to make better purchasing decisions and whose recommendations are based on the customer's preferences and purchasing histories. 
\begin{figure}[t!]
\vspace{-0.5cm}
\centering
 \includegraphics[width=1\linewidth]{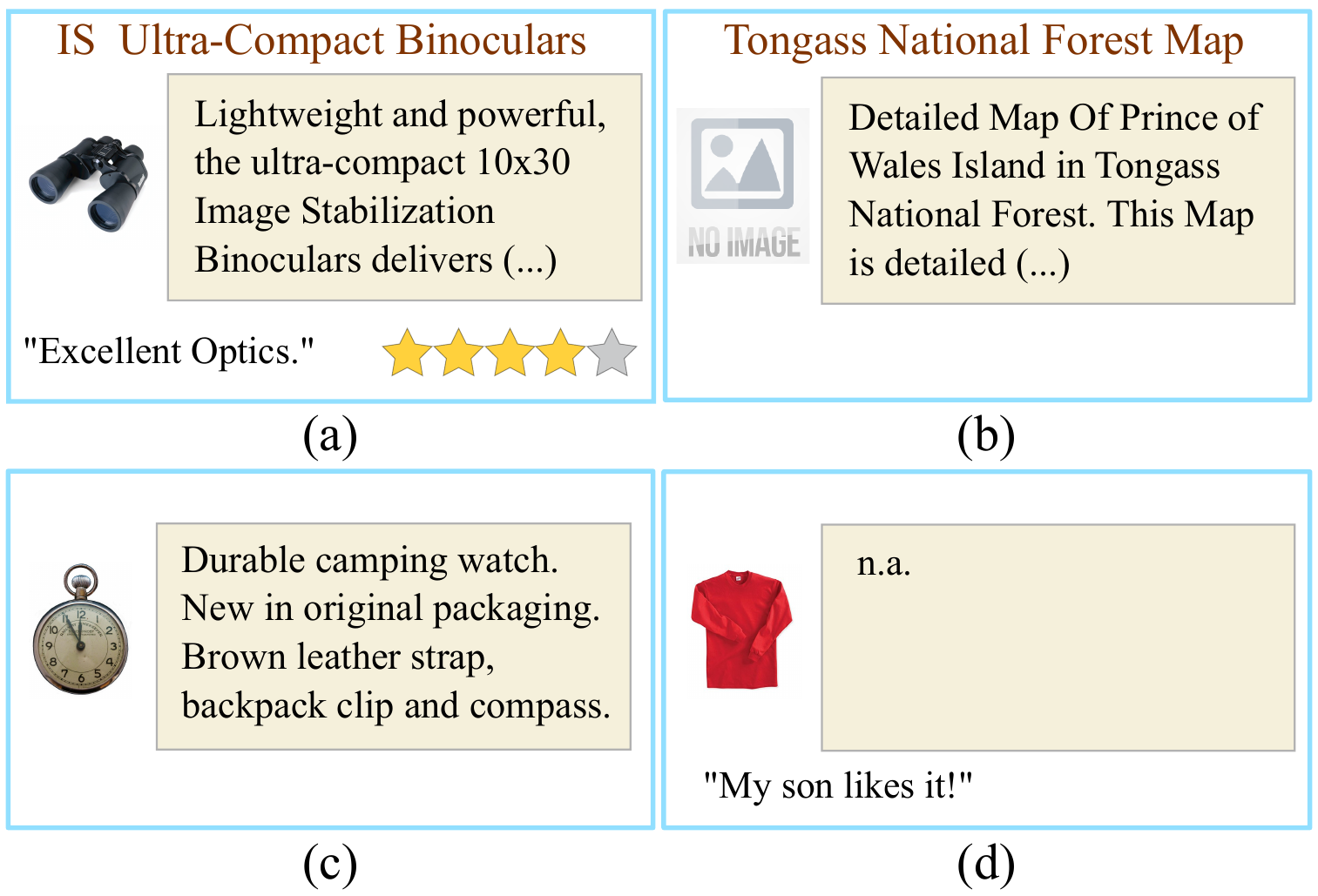}
 \caption{Examples of typical multimodal product data from online retailers: image, title, description, reviews, star ratings. The cold-start problem is present in cases (b) and (c) where neither review text nor ratings are available.}
\label{fig:example_items}
\vskip -0.2in
\end{figure}
Recommender systems can be roughly divided into collaborative filtering (CF)~\cite{koren2009matrix} or content-based filtering (CBF)~\cite{pazzani2007content} methods.  CF-based methods predict the product preference of users based on their previous purchasing and reviewing behavior by computing latent representations of users and products. Standard matrix factorization (MF) and its variants are widely used in CF approaches~\cite{koren2009matrix}. 
While CF-based approaches were demonstrated to perform well in many application domains~\cite{2015rsh}, these methods are based solely on the sparse user-item rating matrix and, therefore, suffer from the so-called \textit{cold-start} problem \cite{schein2002methods, huang2016transferring, wang2017handling} as shown in Figure~\ref{fig:example_items}(b)+(c). For new users without a rating history and newly added products with few or no ratings, the systems fail to generate high-quality personalized recommendations.

Alternatively, CBF approaches incorporate auxiliary modalities/information such as product descriptions, images, and user reviews to alleviate the cold-start problem by leveraging the correlations between multiple data modalities. Unfortunately, a pure CBF method often suffers difficulties in generating a recommendation on incomplete and missing data~\cite{sedhain2015autorec, wang2016collaborative, volkovs2017dropoutnet, abs-1801-10095}. 

In this work, a multimodal imputation framework (LRMM) is proposed to make RS robust to incomplete and missing modalities. First, LRMM learns multimodal correlations~\cite{ngiam2011multimodal, srivastava2012multimodal, wang2016deep, wang2018image} from product images,  product metadata (title+description), and product reviews. We propose modality dropout (\textit{m}-drop) which randomly drops representations of some data modalities. In combination with the modality dropout approach, a sequential autoencoder (\textit{m}-auto) for multi-modal data is trained to reconstruct missing modalities and, at test time, is used to  impute missing modalities through its learned reconstruction function.

Multimodal imputation for recommender systems is a non-trivial issue. (1) Existing RS methods usually assume that all data modalities are available during training and inference. In practice, however, incomplete and missing data modalities are very common. (2) At its core it  addresses the cold-start problem.  In the context of missing modalities, cold-start can be viewed as missing user or item preference information. 

With this paper we make the following contributions:
\begin{itemize} 
\item  For the first time, we introduce multimodal imputation in the context of recommender systems.
\item We reformulate the data-sparsity and cold-start problem when data modalities are missing.
\item We show that the proposed method achieves state-of-the-art results and is competitive with or outperforms existing methods on multiple data sets.
\item We conduct additional  extensive experiments to empirically verify that our approach alleviates the missing data modalities problem. 
\end{itemize}

The rest of paper is structured as follows: Section~\ref{sec:models} introduces our proposed methods. Section~\ref{sec:experiments} describes the experiments and reports on the empirical results. In section~\ref{sec:dis} we discuss the method and its advantages and disadvantages, and in section~\ref{sec:related_work} we discuss related work. Section~\ref{sec:conclusion} concludes this work.

\section{Proposed Methods}
\label{sec:models}

\begin{figure}
\centering
 \includegraphics[width=0.9\linewidth]{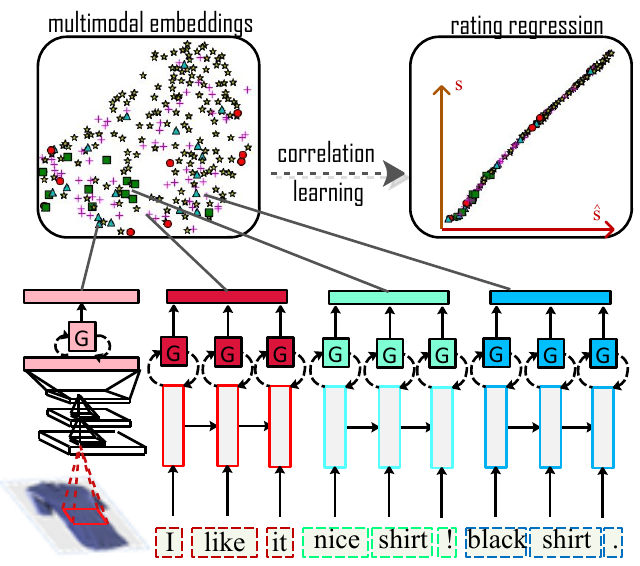}
 \caption{Overview of LRMM. It adopts CNN for visual embeddings (pink part) and three LSTMs for textual embeddings of user review text (red part), item review text (green part) and item meta-data (blue part), respectively. The generative (autoencoder) model is used to reconstruct modality-specific embeddings and impute missing modality. Missing user and item review text lead to user- and item-based cold-start respectively.}
\label{fig:framework}
\vspace{-0.5cm}
\end{figure}

The general framework of LRMM is depicted in Figure \ref{fig:framework}. There are two objectives for LRMM: (1) learning multimodal embeddings that capture inter-modal correlations, complementing missing modalities (Sec. \ref{sec:multimodal_embedding}); (2) learning intra-modal distributions where missing modalities are reconstructed via a missing modality imputation mechanism (sec. \ref{sec:m_dropout} and \ref{sec:msea}).

\subsection{Learning Multimodal Embeddings}
\label{sec:multimodal_embedding}

We denote a user $\bf{u}$ having $k$ review texts as $\mathbf{r}^u$$=$$(r_{o_1}^u, r_{o_2}^u,...,r_{o_k}^u)$  where $r_{o_i}^u$ represents  review text written by $\bf{u}$ for item $o_i$. An item $\bf{o}$ is denoted as $\mathbf{r}^o$$=$$(r_{u_1}^o, r_{u_2}^o,...,r_{u_p}^o)$  where $r_{u_j}^o$ represents the review text written by user $\mathbf{u}_j$ for item $\bf{o}$. Following \citet{zheng2017joint}, to represent each user and item, the reviews of $\bf{u}$ and $\bf{o}$ are concatenated into one review history document: 
\vspace{-0.15cm}
\begin{align}
\mathbf{D}^u=r_{o_1}^u \oplus r_{o_2}^u\oplus,...,\oplus r_{o_k}^u \\
\mathbf{D}^o=r_{u_1}^o \oplus r_{u_2}^o\oplus,...,\oplus r_{u_p}^o
\end{align}
where $\oplus$ is the concatenation operator. Similarly, the metadata of each item $\bf{o}$ can be represented as $\mathbf{D}^m$. For readability, we  use $u,  o,  m,  v$ to denote user, item, metadata, and the image modality, respectively.

For text-based representation learning for user and item, unlike \citet{zheng2017joint} in which CNNs (Convolutional Neural Networks) with Word2Vec \cite{mikolov2013distributed} are employed, our method treats text as sequential data and learns embeddings over word sequences by maximizing the following probabilities: 
\vspace{-0.15cm}
\begin{align}
p(\mathbf{x}_1^g,...,\mathbf{x}_T^g)=\prod\nolimits_{t=1}^{T^g}p(\mathbf{x}_t^g|\mathbf{x}_1^g,...,\mathbf{x}_{t-1}^g)\\
p(\mathbf{x}_t^g|\mathbf{x}_1^g,...,\mathbf{x}_{t-1}^g)=p(\mathbf{x}_t^g|\mathbf{e}_t^g)\\
\mathbf{e}_t^g=\mathcal{M}^g(\mathbf{e}_{t-1}^g,\mathbf{x}_t^g; \mathbf{\Theta}^g)
\label{equ:recu}
\end{align} 
where $\mathcal{M}^g$, $g \in \{u,o,m\}$ is a recurrent model and $(\mathbf{x}_1^g,...,\mathbf{x}_T^g)$ is the word sequence of either review or metadata text, each $\mathbf{x}_t^g\in \mathcal{V}$ and $\mathcal{V}$ is a vocabulary set.  $T^g$ is the length of input and output sequence and $\mathbf{e}_t^g$ is the hidden state computed from the corresponding LSTM (Long Short Term Memory) \cite{hochreiter1997long} by: 
\begin{align}
\label{equ:input}
&
\begin{pmatrix}\mathbf{i}_t
\\ \mathbf{f}_t
\\ \mathbf{o}_t
\\ \mathbf{g}_t
\end{pmatrix}= \begin{pmatrix}
\text{sigm}
\\ \text{sigm}
\\ \text{sigm}
\\ \text{tanh}
\end{pmatrix}\mathbf{W}
\begin{pmatrix} \mathbf{x}_t
\\ \mathbf{h}_{t-1}
\end{pmatrix}\\
&
\label{equ:cell}
\mathbf{c}_t=\mathbf{f}_t\odot\mathbf{c}_{t-1}+\mathbf{i}_t\odot\mathbf{g}_t\\
&
\mathbf{h}_t=\mathbf{o}_t\odot\tanh(\mathbf{c}_t)
\end{align}
where $\mathbf{i}_t$, $\mathbf{f}_t$ and $\mathbf{o}_t$ are input, forget and output gate respectively, $\mathbf{c}_t$ is memory cell, $\mathbf{h}_t$ is the hidden output that we used for computing user or item embedding $\mathbf{e}^g, ~g\in \{u,o,m\}$.

As we treat each text document $\mathbf{D}^g$ as a word sequence of length $T^g$, we adopt average pooling on word embeddings for each modality to obtain document-level representations: 
\begin{equation}
\mathbf{e}^g=\frac{\sum\nolimits_{t\in T^g, ~g\in \{u,o,m\}}\mathbf{e}_t^g}{T^g}
\end{equation}
Visual embeddings $\mathbf{e}^v$ are extracted with a pre-trained CNN and transformed by a function $f$
\begin{equation}
\mathbf{e}^v=f(\mbox{CNN}(I, \mathbf{\Theta}_c);\mathbf{\Theta}_f),
\end{equation}
where $\mathbf{\Theta}_f \in \mathbb{R}^{4096\times d}$ to ensure $\mathbf{e}^v$ has same dimension as the user $\mathbf{e}^u$, item  $\mathbf{e}^o$, and metadata embedding $\mathbf{e}^m$. The multimodal joint embedding then can be learned by a shared layer and used for making a prediction:
\begin{equation}
\mathbf{\hat{s}}=f_s(\mathbf{W}_s(\mathbf{e}^u \oplus \mathbf{e}^o \oplus \mathbf{e}^m \oplus \mathbf{e}^v)+\mathbf{b}_s)
\label{equ:concate}
\end{equation} 
where $f_s:\mathbb{R}^{4\times d}\rightarrow\mathbb{R}^1$, parameterized with $\mathbf{W}_s$ and $\mathbf{b}_s$, is a scoring function to map the multimodal joint embedding to a rating score. 

\begin{figure}[!t]
\centering
 \includegraphics[width=1\linewidth]{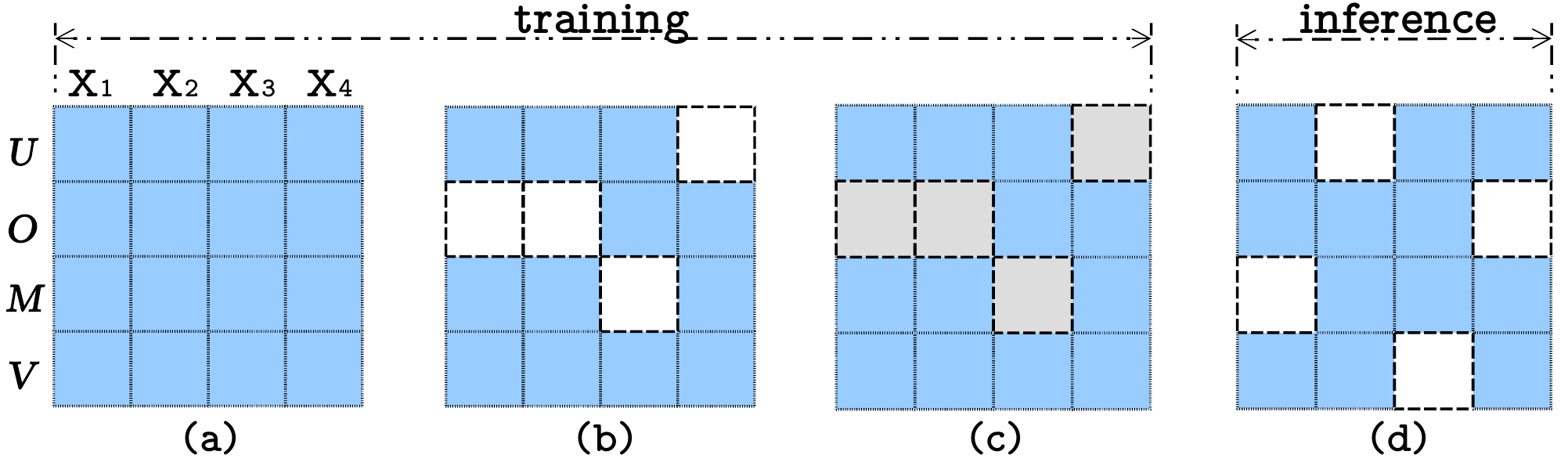}
 \caption{Missing modality imputation. (a) Full training data, (b) \textit{m}-drop randomly drops modalities, (c) \textit{m}-auto learns to reconstruct missing data based on existing data. (d) Inference with missing modalities. Dropping user and item view is equivalent to learning models being able to address cold-start problem.}
\label{fig:technique}
\vskip -0.1in
\label{fig:drop}
\end{figure}

\subsection{Modality Dropout}
\label{sec:m_dropout}
Modality dropout (\textit{m}-drop) is designed to remove a data modality during training according to some parametric distribution.  This is motivated by dropout \cite{srivastava2014dropout} which randomly masks hidden layer activations to zero to increase the generalization capability of the underlying model. More formally, \textit{m}-drop changes the original feed-forward equation: 
\begin{equation}
\mathbf{Z}^{(L+1)}=\varphi(\mathbf{W}^{(L+1))}\mathbf{X}^{(L)}+\mathbf{b}^{(L+1)})
\end{equation} 
being able to randomly drop modality by:
\begin{align}
&r^{(L)} \sim \text{Bernoulli}~(p_m) \\
&k^{(L)} \sim \text{Bernoulli}~(1-1/n_m)\\
&\mathbf{\widetilde{X}}^{(L)}=(\mathbf{X}^{(L)}\odot \mathbf{r}^{(L)})\odot \mathbf{k}^{(L)}\\
&\mathbf{Z}^{(L+1)}=\varphi(\mathbf{W}^{(L+1))}\mathbf{\widetilde{X}}^{(L)}+\mathbf{b}^{(L+1)})
\end{align} 
where each sample $\mathbf{X}_1={x_1, ..., x_{n_m}}$ and $n_m$ is the number of modalities. $\mathbf{r}^{(L)}$ is a vector of independent Bernoulli random variables each of which has probability $p_m$ of being 1. $\mathbf{k}^{(L)}$ is a vector of independent variables which indicate the dropout on modality with a given probability. $\varphi(\cdot)$ is an activation function. 

Figure \ref{fig:technique} (a-b) shows how \textit{m}-drop works. Note the differences between modality dropout (\textit{m}-drop) and original dropout: (1) \textit{m}-drop targets specifically the multimodal scenario where some modalities are completely missing; and (2) \textit{m}-drop is performed on the input layer ($L\equiv 0$).

\subsection{Mutlimodal Sequential Autoencoder}
\label{sec:msea}
The autoencoder has been used in prior work \cite{sedhain2015autorec,strub2016hybrid} to reconstruct missing elements (mostly ratings) in recommender systems.  This is equivalent to the case of missing at random (MAR). For MAR, it is rare to have a continuous large block of missing entries \cite{tran2017missing}. Differently, in recommending with missing modality, the missing entries typically occur in a \textit{large continuous block}. For instance, an extreme case is the absence of all item reviews and ratings (data sparsity is 100\%, leading to the so-called item cold-start problem).
Existing methods~\cite{lee2001algorithms,koren2008factorization,marlin2004modeling,wang2011collaborative,mcauley2013hidden,Li:2017:NRR,zheng2017joint} have difficulties when entire data modalities are missing during the \textit{training} and/or \textit{inference} stages. 

To address this limitation, we propose a multimodal sequence autoencoder (\textit{m}-auto) to impute textual sequential embeddings and visual embeddings for the missing modalities. Modality-specific autoencoders are placed between the modality-specific encoders (i.e., CNN and LSTMs) and the shared layer (equation \ref{equ:concate}). The reconstruction layers, therefore, can capture the inter-modal and intra-modal correlations. More formally, for each data modality $g\in \{u,o,m,v\}$, the modality-specific encoder is given as
\begin{equation}
\mathbf{e}_{hid}^g=\text{sigm}(\mathbf{W}_{vh}^g\mathbf{e}_{in}^g+\mathbf{b}_{vh}^g)
\label{equ:ae_encoder}
\end{equation} 
and the modality-decoder is given as
\begin{align}
\mathbf{e}_{recon}^g=\frac{1}{T^g}\sum_{t\in T^g}\text{sigm}(\mathbf{W}_{hv}^g\mathbf{e}_{hid}^g+\mathbf{b}_{hv}^g)
\label{equ:ae_decoder}
\end{align} 
where $\mathbf{W}_{vh}\in\mathbb{R}^{d \times d_h}$ and $\mathbf{W}_{hv} \in\mathbb{R}^{d_h \times d} $ are weights, $\mathbf{b}_{vh}$, $\mathbf{b}_{hv}$ are biases receptively for visible-to-hidden, and hidden-to-visible layers. $\mathbf{e}_{in}^g$,  $\mathbf{e}_{hid}^g$ present the original, hidden word-level embeddings, and $\mathbf{e}_{recon}^g$ is the reconstructed document-level embeddings. The $\mathbf{e}^g$ is a modality-specific embedding.
 
\textit{m}-auto is different from previous reconstruction models\cite{sedhain2015autorec,strub2016hybrid} in that its reconstructions are based on inter-modal and intra-modal correlations in the context of multimodal learning.

\subsection{Model Optimization}
The optimization of the network is formulated as a regression problem by minimizing the mean squared error (MSE) loss $\mathcal{L}_{reg}$:
\begin{equation}
\mathcal{L}_{reg}=\frac{1}{\mathcal{|D|}}\sum_{(u, o, m, v)\in \mathcal{D}}(\mathbf{\hat{s}}-\mathbf{s})^2+\lambda \parallel \mathbf{\Theta}_r  \parallel_2
\label{}
\end{equation}
where $\mathbf{\hat{s}}$ and $\mathbf{s}$ are the predicted and truth rating scores. $\mathcal{|D|}$ is dataset size , $\lambda$ is weight decay parameter and $ \mathbf{\Theta}_r$ is regression model parameters.
To constrain the representations to be compact in reconstruction, a penalty term is utilized
\begin{equation} 
\mathcal{H}=\sum_{i=1}^{h_n}\log\frac{\rho}{\hat{\rho}_i}+(1-\rho)\log\frac{1-\rho}{1-\hat{\rho}_i}
\end{equation}
where $\rho$ and $\hat{\rho}$ are sparsity parameters and average activation of hidden unit $i$, $h_n$ is the number of hidden units. The reconstruction loss for each modality is now
\begin{equation}
\begin{split}
\mathcal{L}_{recon}^g= &\frac{1}{\mathcal{|D|}}\sum_{g\in \{u,o,m,v\}}\left \| \mathbf{e}_{recon}^{g}-\mathbf{e}_{in}^{g}\right \|^2 \\
&+ \lambda_{\rho}\sum_{g\in \{u,o,m,v\}}\mathcal{H}^g
\end{split}
\end{equation}
where $\lambda_{\rho}$ is a sparsity regularization term. 
The objective of the entire model is then
\begin{align}
\begin{split}
\mathcal{L}=\alpha \mathcal{L}_{reg}+\beta \sum_{g\in \{u,o,m,v\}}\mathcal{L}_{recon}^g
\end{split}
\end{align} 
where $\alpha$ and $\beta$ are learnable parameters. The model is learned in an end-to-end fashion through back-prorogation~\cite{lecun1989backpropagation}.

\section{Experiments}
\label{sec:experiments}
This section evaluates LRMM on rating prediction tasks with real-world datasets. We firstly compare LRMM with recent methods (sec. \ref{sec:compare}), then we empirically show the effectiveness of LRMM in alleviating the cold-start, the incomplete/missing data, and the data sparsity problem (sec. \ref{sec:cold}-\ref{sec:transfer}).

\subsection{Datasets and Evaluation Metrics}

We conducted experiments on the Amazon dataset~\cite{mcauley2015image, he2016ups}\footnote{\url{http://jmcauley.ucsd.edu/data/amazon/}}, which is widely used for the study of recommender systems. It consists of different modalities such as text, image, and numerical data. We used 4 out of 21 categories: Sports and Outdoors (S\&O), Health and Personal Care (H\&P), Movies and TV, Electronics. Some statistics of the datasets are listed in Table \ref{tab:dataset}. We randomly split each dataset into 80\% training, 10\% validation, and 10\% test data. Each input instance consists of four parts $\mathbf{x}^{(i)}$$=$$(
\mathbf{x}^{(i)}_u,
\mathbf{x}^{(i)}_o,
\mathbf{x}^{(i)}_m,
\mathbf{x}^{(i)}_v)$,
where $\mathbf{x}^{(i)}_u$ and $\mathbf{x}^{(i)}_i$ are the concatenated reviews of users and items in the training data. 
$\mathcal{V}$ is the vocabulary that was built based on reviews and metadata on the training data. Words with an absolute frequency of at least $20$ are included in the vocabulary.

\begin{table}
\centering
\caption{Datasets}
\label{tab:dataset}
\begin{tabular}{p{1.1cm}|p{1.05cm}p{1.05cm}p{1.05cm}p{1.4cm}}
\hline
 Dataset     & S\&O    &  H\&P    &  Movie   &  Electronics \\ \hline\hline
Users   & 35494  & 38599  & 111149 & 192220     \\
Items   & 16415  & 17909  & 27019   & 59782       \\
Samples & 272453 & 336769 & 974582 & 1614105   \\
$\mathcal{|V|}$   & 42095  & 47476  & 160117 & 198598  \\  \hline
\end{tabular}
\vskip -0.1in
\end{table}

To evaluate the proposed models on the task of rating prediction, we employed two metrics, namely, Root Mean Square Error (RMSE)
\begin{equation}
RMSE=\sqrt{\frac{1}{\mathcal{|D|}} \sum_{(u, o, m, v) \in \mathcal{D}}(\mathbf{\hat{s}}_{i,j}-\mathbf{s}_{i,j})^2}
\end{equation} and Mean Absolute Error (MAE)
\begin{equation}
MAE=\frac{1}{\mathcal{|D|}} \sum_{(u, o, m, v) \in \mathcal{D}}\left | \mathbf{\hat{s}}_{i,j}-\mathbf{s}_{i,j} \right |
\end{equation}
where $\mathbf{\hat{s}}_{i,j}$ and $\mathbf{s}_{i,j}$ represent the predicted rating score and ground truth rating score that user $i$ gave to item $j$.

\begin{table*}[!t]
\caption{\label{tab:comparison_S_H}Comparison on datasets with the baselines.  `+F': tested with all modalities(U,O,M,V), `-X': dropping one modality, `-U' and `-O': user and item cold-start scenario.}
\centering
\begin{center}
\vskip -0.1in
\begin{tabular}{l|cccccccc}
\hline
  Dataset    & \multicolumn{2}{c}{ S\&O} & \multicolumn{2}{c}{ H\&P}  & \multicolumn{2}{c}{\ Movie} & \multicolumn{2}{c}{ Electronics} \\ 
 Models     &   RMSE        &  MAE         &  RMSE        &  MAE  &   RMSE        &  MAE         &  RMSE        &  MAE        \\ \hline \hline
Offset  & 0.979      & 0.769      & 1.247      & 0.882   & 1.389      & 0.933      & 1.401      & 0.928   \\ 
NMF   & 0.948      & 0.671      & 1.059      & 0.761     & 1.135      & 0.794      & 1.297      & 0.904  \\ 
SVD++ & 0.922      & 0.669      & 1.026     & 0.760      & 1.049      & 0.745      & 1.194      & 0.847 \\ 
URP  & - & - & - &- &   1.006    & 0.764       & 1.126      & 0.860     \\ 
RMR  & - & - & - &-  &  1.005     & 0.741      & 1.123      & 0.822     \\ 
HFT & 0.924    & 0.659      & 1.040     & 0.757     &  0.997     & 0.735      & 1.110      & 0.807   \\ 
DeepCoNN & 0.943    & 0.667      &  1.045    & 0.746    & 1.014     &    0.743            &  1.109               & 0.797  \\
NRT & - & - & - &-  &  0.985      & \bf 0.702      &   1.107    &  0.806    \\  \hline \hline 
LRMM(+F)  & \bf 0.886      & \bf 0.624      & \bf 0.988      & \bf 0.708   & \bf 0.983      & 0.716      &  \bf 1.052     & \bf 0.766       \\ 
LRMM(-U)   & 0.936      & 0.719      & 1.058      & 0.782     & 1.086     & 0.821      &  1.138     &  0.900  \\ 
LRMM(-O)   & 0.931     & 0.680      & 1.039      & 0.805      & 1.074      & 0.855      &   1.101    &   0.864  \\ 
LRMM(-M)   & 0.887      & 0.625      & 0.989      & 0.710  & 0.991      & 0.725     &  1.053     &   0.766     \\ 
LRMM(-V)   & 0.886      & 0.624      & 0.989      & 0.708      & 0.991     & 0.725      &   1.052    &   0.766 \\ \hline
\end{tabular}
\end{center}
\vskip -0.2in
\end{table*}

\subsection{Baselines and Competing Methods}
\label{sec:comp_method}
We compare our models with several baselines\footnotemark. The baselines can be categorized into three groups.
\begin{itemize}
\item[(1)] Matrix factorization: \textit{NMF} \cite{lee2001algorithms} and \textit{SVD++} \cite{koren2008factorization}. 
\item[(2)] Topic model methods: \textit{URP} \cite{marlin2004modeling}, \textit{CTR}~\cite{wang2011collaborative}, \textit{HFT}~\cite{mcauley2013hidden} and \textit{RMR}~\cite{Ling2014RMR}. 
\item[(3)] Deep learning models: \textit{NRT}~\cite{Li:2017:NRR} and \textit{DeepCoNN}~\cite{zheng2017joint}, which are current state-of-the-art approaches. 
\end{itemize}
We also include a naive method---\textit{Offset} \cite{mcauley2013hidden} which simply takes the average across all training ratings.

\footnotetext{To make a fair comparison, implemented baselines are trained with grid search (for NMF and SVD++, regularization [0.0001, 0.0005, 0.001], learning rate [0.0005, 0.001, 0.005, 0.01]. For HFT, regularization [0.0001, 0.001, 0.01, 0.1, 1], lambda [0.1, 0.25, 0.5, 1]). For DeepCoNN, we use the suggested default parameters. The best scores are reported.}

\subsection{Implementation}
We implemented LRMM with Theano\footnote{\url{http://www.deeplearning.net/software/theano/}}. The weights for the non-recurrent layer were initialized by drawing from the interval $\left [ -\sqrt{\frac{6}{N_{in}+N_{out}}},\sqrt{\frac{6}{N_{in}+N_{out}}} \right ]$ ($N$ is the number of units) uniformly at random. We used 1024 hidden units for the autoencoder. The LSTMs have 256 hidden units and the internal weights $\mathbf{W}$ are orthogonally initialized~\cite{saxe2013exact}. We used a batch size of 256, $\lambda=0.0001$, sparsity parameter $\rho=0.05$, $\lambda_{\rho}=0.01$, an initial learning rate of 0.0001 and a dropout rate of 0.5 after the recurrent layer. The models were optimized with ADADELTA \cite{zeiler2012adadelta}. The length of the user, item and meta-data document $\mathbf{D}^u$, $\mathbf{D}^o$, and $\mathbf{D}^m_o$ were fixed to $L=100$. We truncated documents with more than 100 words. The image features are extracted from the first fully-connected layer of CNN on ImageNet \cite{russakovsky2015imagenet}.

We implemented NMF and SVD++ with the SurPrise package\footnote{\url{http://surpriselib.com/}}. Offset and HFT were implemented by modifying authors' implementation\footnote{\url{http://cseweb.ucsd.edu/~jmcauley/code/code_RecSys13.tar.gz}}. For DeepCoNN, we adapted the implementation from \cite{Chong2018}\footnote{\url{https://github.com/chenchongthu/DeepCoNN}}. The numbers of other methods are taken from \citet{Li:2017:NRR}.  

\subsection{Compare with State-of-the-art}
\label{sec:compare}

First, we compare LRMM with state-of-the-art methods listed in Sec. \ref{sec:comp_method}. In this setting, LRMM is trained with all data modalities and tested with different missing modality regimes. Table~\ref{tab:comparison_S_H} lists the results on the four datasets. By leveraging multimodal correlations, LRMM significantly outperforms MF-based models (i.e. NMF, SVD++) and topic-based methods (i.e., URP, CTR, RMR, and HFT).  LRMM also outperforms recent deep learning models (i.e., NRT, DeepCoNN) with respect to almost all metrics.

LRMM is the only method with a robust performance for the cold-start recommendation problem where user review or item review texts are removed. While the cold-start recommendation is more challenging, LRMM(-U) and LRMM(-O) are still able to achieve a similar performance to the baselines in the standard recommendation setting. For example, RMSE 1.101 (LRMM(-O)) to 1.107 (NRT) on Electronics, MAE 0.680 (LRMM(-O)) to 0.667 (DeepCoNN)on S\&O.  We conjecture that the cross-modality dependencies  \cite{srivastava2012multimodal} make LRMM more robust when modalities are missing. 
Table~\ref{fig:example_rating} lists some randomly selected rating predictions. Similar to Table~\ref{tab:comparison_S_H}, missing user (-U) and item (-O) preference significantly deteriorates the performance.  

\begin{figure}
	\center
        \includegraphics[width=0.45\textwidth]{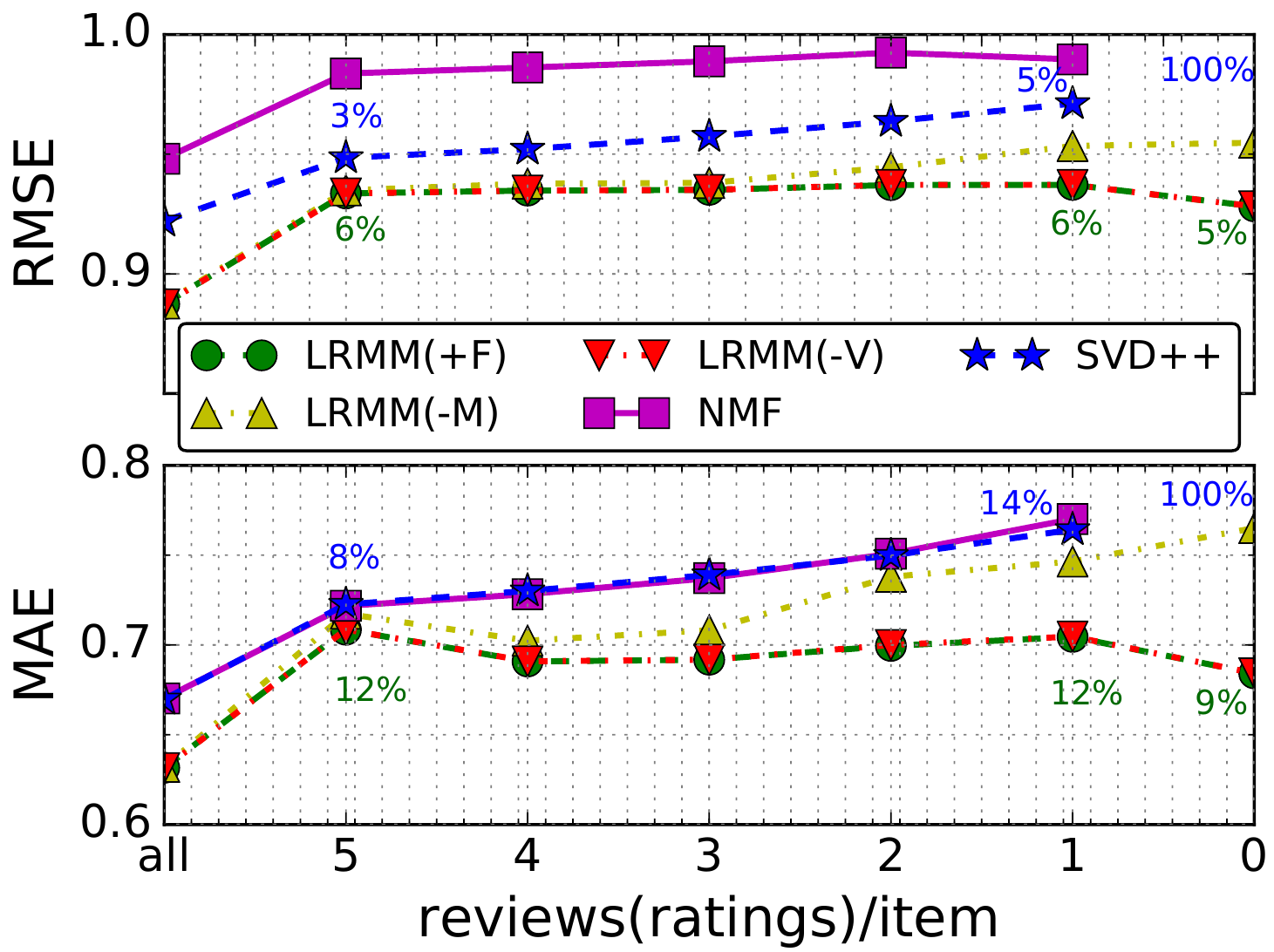}
        \caption{Performance with reduced reviews (ratings) on S\&O dataset. (\% : performance drops for SVD++ (in blue) and LRMM(+F) (in green))}
 	\label{fig:rmse_mae_reduce}
\vskip -0.in
 \end{figure}
 
\subsection{Cold-Start Recommendation}
\label{sec:cold}

Prior work \cite{mcauley2013hidden, zhang2017deep} has considered users (items) with sparse preference information as the cold-start problem (e.g., Figure \ref{fig:example_items}(d)), that is, where there is still some information available. In practice, preference information could be missing in larger quantities or even be entirely absent (e.g., Figure \ref{fig:example_items}(b-c)). In this situation, the aforementioned methods are not applicable as they require some data to work with. In this experiment, we examine how LRMM leverages modality correlations to alleviate the data sparsity problem when training data becomes even sparser. To this end, we train models for the item cold-start problem by reducing the number of reviews (for LRMM) and ratings (for NMF and SVD++) of each item in the training set.  

Figure \ref{fig:rmse_mae_reduce} demonstrates the robustness of LRMM when the training data becomes more sparse.  Note that NMF and SVD++ fail to train models when there is no ratings data available. In contrast, LRMM is trained by leveraging item images and metadata even if item reviews are completely missing for a product. The average number of reviews per item on this dataset is 16.7. Reducing the number of ratings to 5 severely degrades the performance of NMF, SVD++, and LRMM.  However, LRMM remains rather stable in maintaining good performance when considering the performance degradation at 5, 1, and 0 reviews (ratings), respectively. One interesting observation is that, with a reduced number of reviews, the product metadata plays a more and more important role in maintaining the performance: LRMM(-V) is close to LRMM(+F) in Figure \ref{fig:rmse_mae_reduce} while the gap between LRMM(-M) and LRMM(+F) is large.

\subsection{Missing Modality Imputation}
\label{sec:missing}
\begin{table}[]
\centering
\caption{\label{tab:mdrop}The performance of training with missing modality imputation.}
\begin{tabular}{l|cccc}
\hline
Dataset & \multicolumn{2}{c}{S\&O} & \multicolumn{2}{c}{H\&P} \\ 
Models  & RMSE        & MAE         & RMSE        & MAE         \\  \hline\hline
LRMM(+F)  & 0.997       & 0.790       & 1.131       & 0.912       \\ 
LRMM(-U)  & 0.998       & 0.795       & 1.132       & 0.914       \\ 
LRMM(-O)  & 0.999       & 0.796       & 1.133       & 0.917       \\ 
LRMM(-M)  & 0.998       & 0.797       & 1.133       & 0.913       \\ 
LRMM(-V)  & 0.997       & 0.791       & 1.132       & 0.913       \\ \hline
\end{tabular}
\vskip -0.in
\end{table}

\begin{figure}[t!]
\vskip -0.in
    \centering
    \begin{subfigure}[t]{0.11\textwidth}
        \centering
        \includegraphics[height=1.1in]{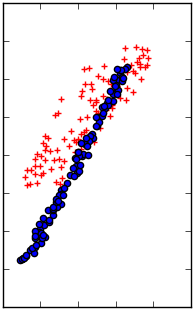}
        \caption{$\mathbf{e}^{v}$}
    \end{subfigure}%
    \begin{subfigure}[t]{0.11\textwidth}
        \centering
        \includegraphics[height=1.1in]{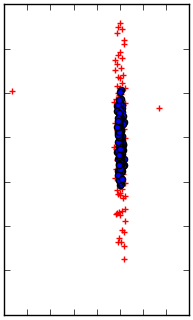}
        \caption{$\mathbf{e}^{u}$}
    \end{subfigure}
    \begin{subfigure}[t]{0.11\textwidth}
        \centering
        \includegraphics[height=1.1in]{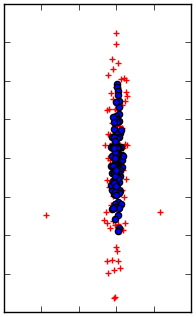}
        \caption{$\mathbf{e}^{o}$}
    \end{subfigure}
    \begin{subfigure}[t]{0.11\textwidth}
        \centering
        \includegraphics[height=1.1in]{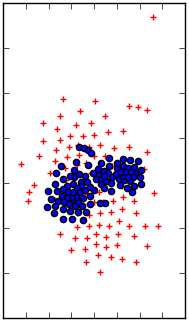}
        \caption{$\mathbf{e}^{m}$}
    \end{subfigure}
    \caption{Visualization of embeddings (blue) and reconstructed embeddings (red)}
    \label{fig:recon_embeddings}
\vskip -0.in
\end{figure}
The proposed \textit{m}-drop and \textit{m}-auto methods allow LRMM to be more robust to missing data modalities. Table~\ref{tab:mdrop} lists the results of training LRMM with missing data modalities for the modality dropout ratio $p_m= 0.5$ on the S\&O and H\&P datasets, respectively. Both RMSE and MAE of LRMM deteriorate but are still comparable to the MF-based approaches NMF and SVD++. However, the proposed method LRMM is robust to missing data in both training and inference stages, a problem rarely addressed by existing approaches. 
In Figure~\ref{fig:recon_embeddings}, we visualized the modality-specific embeddings and their reconstructed embeddings of 100 randomly selected samples with t-SNE~\cite{maaten2008visualizing}. The plots suggest that it is more challenging to reconstruct item metadata and image embeddings as compared to the user or item embeddings. One possible explanation is that some selected metadata contains noisy data (e.g., ``\textit{ISBN - 9780963235985}'', ``\textit{size: 24 $\times$46}'' and ``\textit{Dimensions: 15W $\times$  22H}'') for which visual data is more diverse. This would increase the difficulty of incorporating visual data into the embeddings.

\subsection{The Effect of Text Length}
\label{sec:length}

\begin{figure}[t]
\vskip -0.in
    \centering
    \begin{subfigure}[t]{0.245\textwidth}
        \includegraphics[ width=0.995\textwidth]{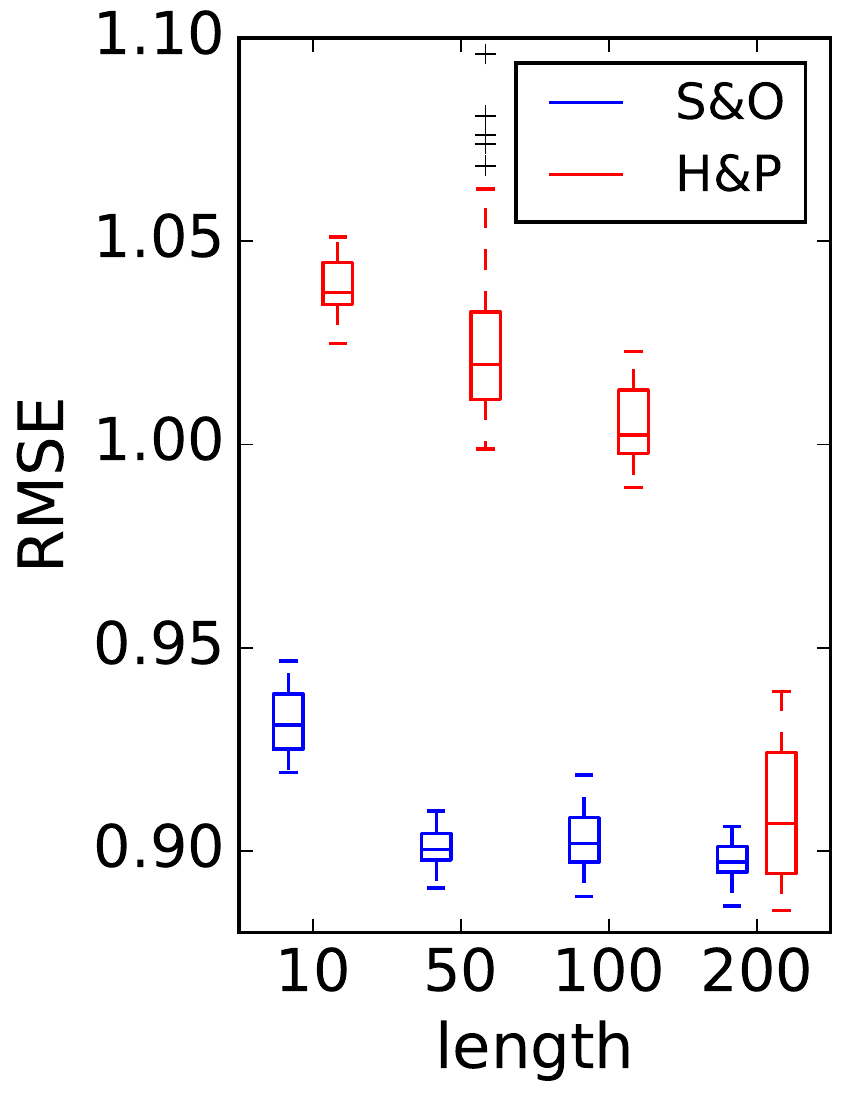}
    \end{subfigure}%
    \begin{subfigure}[t]{0.245\textwidth}
        \includegraphics[width=0.995\textwidth]{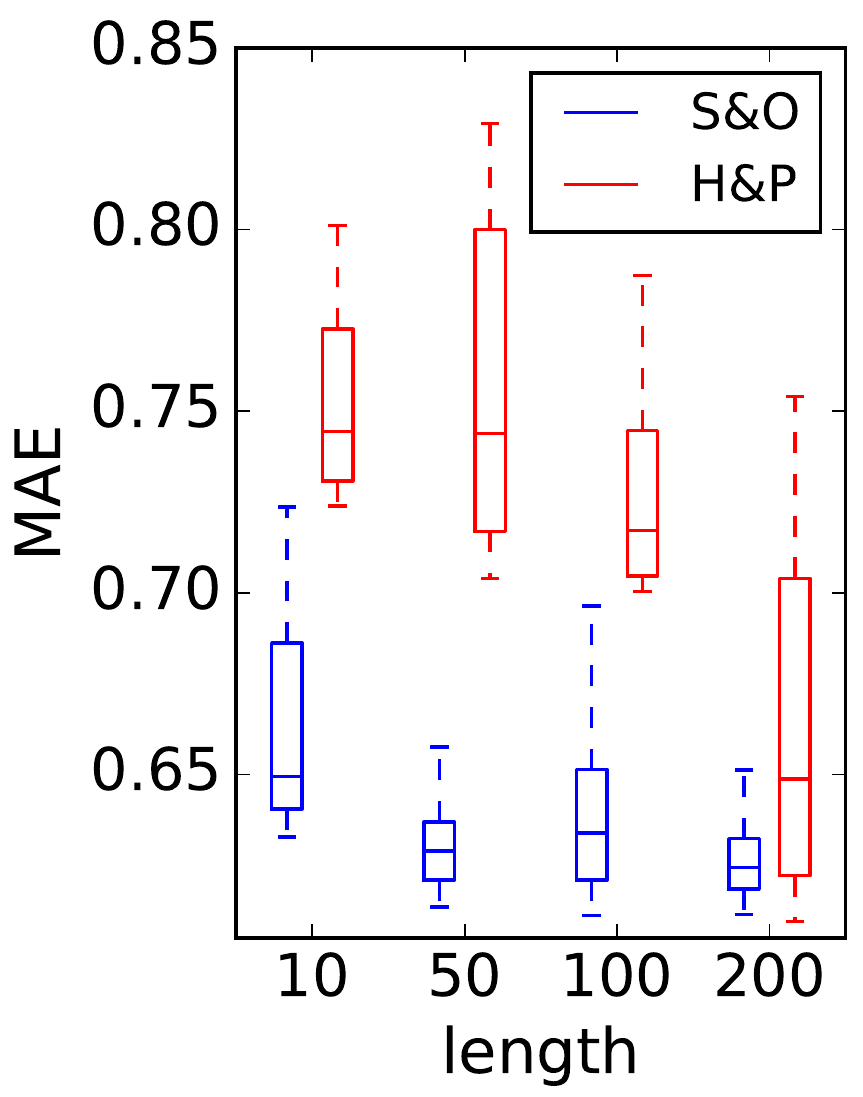}
    \end{subfigure}
    \caption{RMSE and MAE with varied text lengths on S\&O and H\&P datasets.}
    \label{fig:rmse_mae_lengths}
\vskip -0.in
\end{figure}

To alleviate the data sparsity problem, existing work \cite{mcauley2013hidden, zhang2017deep} concatenates review texts and utilizes topic modeling (e.g. HFT) or CNNs combined with Word2Vec (e.g. DeepCoNN) to learn user or item embeddings. Differently, LRMM treats the concatenated reviews as sequential data and learns sequence embeddings with RNNs.  In this experiment, we show that learning sequential embedding is beneficial on sparse data because it is unnecessary to exploit all reviews so as to reach good performance.   
Figure \ref{fig:rmse_mae_lengths} shows the performance of LRMM with varied word sequence lengths. In general, sequence embeddings learned with larger length achieve better performance. Note that, by considering a certain amount of words (e.g. L=50), LRMM is able to achieve a result as good as accounting more words (e.g. L=100 or 200). 
Although this is dataset-dependent to some degree, e.g., LRMM (L=200) improves RMSE and MAE in a certain margin as compared to L=100 on the H\&P data, it demonstrates the superiority of sequential user or item embeddings as compared to topic and CNN+Word2Vec embeddings on more sparse data as shown in Table~\ref{tab:comparison_S_H}. 

\subsection{Cross-Domain Adaptation}
\label{sec:transfer}

To consider an even more challenging situation we explore cases where the full training set is missing. Inspired by the recent success of domain adaptation (DA)~\cite{csurka2017domain}, a special form of transfer learning~\cite{pan2010survey,weiss2016survey}, we perform the recommendation task on the target domain  test set $\mathcal{D}^t_{test}$ (e.g., ``Sport'') but with the model $\mathcal{C}$ trained on a different domain  training set $\mathcal{D}^s_{train}$ (e.g. ``Movie''). This is achieved by extracting the multimodal embeddings on the source domain and by performing prediction on the target domain. Table \ref{tab:multi_domain} shows the performance of LRMM when performing adaptation from larger datasets to smaller datasets. Although the performance is not as good as on $\mathcal{D}^s_{test}$, LRMM is still able to obtain decent results even without using training data $\mathcal{D}^t_{train}$. Table~\ref{fig:mda} shows some example rating predictions on DA for different categories of products. It demonstrates the strong generalization capability of DA from one product category to another. 

\begin{table}[t!]
\centering
\caption{Cross-Domain Adaptation with LRMM}
\label{tab:multi_domain}
\begin{tabular}{l|p{0.605cm}p{0.605cm}p{0.605cm}p{0.605cm}p{0.605cm}} \hline
$\mathcal{D}^s\rightarrow\mathcal{D}^t$ & \bf +F & \bf -U & \bf -O & \bf -M & \bf -V \\ \hline \hline
Movie$\rightarrow$S\&O   & 1.061  & 1.013  & 1.071  & 1.061  & 1.062  \\ 
Movie$\rightarrow$H\&P   & 1.190  & 1.140  & 1.170  & 1.190  & 1.190  \\ 
Elect.$\rightarrow$S\&O  & 1.072  & 1.012  & 1.088  & 1.073  & 1.073  \\ 
Elect.$\rightarrow $H\&P & 1.191  & 1.137  & 1.180  & 1.191  & 1.192  \\ \hline
\end{tabular}
\vskip -0.in
\end{table}

\begin{table}[t!]
\footnotesize
\centering
\caption{Exemplary rating prediction on S\&O datatset. `T' means true ratings, the best prediction is in blue, the worst prediction is in red.}
\label{fig:example_rating}
\begin{tabular}{p{1.5cm}|c|ccccc} \hline
Item image & \bf T & \bf +F & \bf -U & \bf -O & \bf -M & \bf -V \\ \hline \hline
\begin{minipage}{0.1\textwidth}
\centering
\includegraphics[width=0.4\linewidth,  height=0.02\textheight]{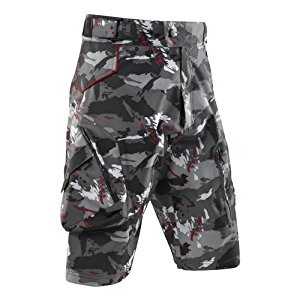}
\end{minipage} & 3  &  3.18  & \color{red}\textbf{3.97}   & 3.48   & \color{blue}\textbf{3.03}   & 3.28  \\ 

 \begin{minipage}{0.1\textwidth}
\centering
\includegraphics[width=0.4\linewidth,  height=0.02\textheight]{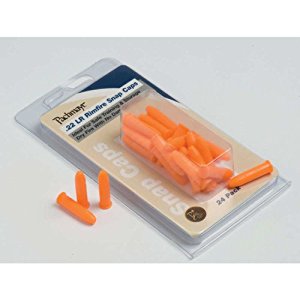}
\end{minipage} & 3  &  3.36 & \color{red}\textbf{4.07}   & 3.5   & 3.33   & \color{blue}\textbf{3.27}  \\ 

\begin{minipage}{0.1\textwidth}
\centering
\includegraphics[width=0.4\linewidth, height=0.02\textheight]{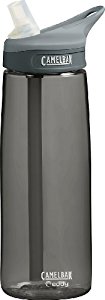}
\end{minipage} &  5 &  \color{blue}\textbf{4.63}  & 4.50   & \color{red}\textbf{4.36}   & 4.60   &\color{blue}\textbf{4.63}   \\ 

\begin{minipage}{0.1\textwidth}
\centering
\includegraphics[width=0.4\linewidth, height=0.02\textheight]{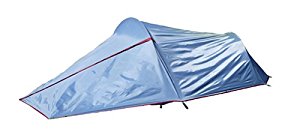}
\end{minipage} &  3 &  \color{blue}\textbf{3.11}  & \color{red}\textbf{3.77}   & 3.49   & 3.44   & 3.57   \\ 

\begin{minipage}{0.1\textwidth}
\centering
\includegraphics[width=0.4\linewidth, height=0.02\textheight]{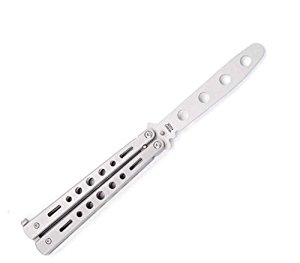}
\end{minipage} &  4 &  \color{blue}\textbf{4.00}  & \color{red}\textbf{4.31}   & 3.87   & 3.92   & 4.02   \\ 
 \hline
\end{tabular}
\vskip -0.0in
\end{table}

\begin{table}[t!]
\footnotesize
\centering
\caption{Examples on H\&P datatset with domain adaptation. The model is trained on Movie dataset.}
\label{fig:mda}
\begin{tabular}{p{1.5cm}|c|ccccc} \hline
Item image& \bf T & \bf +F & \bf -U & \bf -O & \bf -M & \bf -V \\ \hline \hline
\begin{minipage}{0.1\textwidth}
\centering
\includegraphics[width=0.4\linewidth,  height=0.02\textheight]{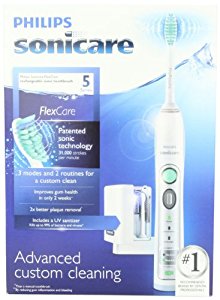}
\end{minipage} & 4  &   \color{blue}\textbf{4.01}  &4.18   &  \color{red}\textbf{3.70}   &4.04   & 4.05  \\ 

 \begin{minipage}{0.1\textwidth}
\centering
\includegraphics[width=0.4\linewidth,  height=0.02\textheight]{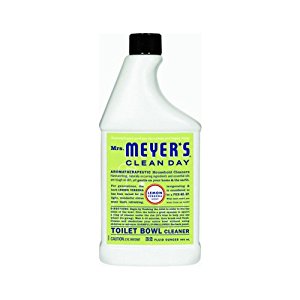}
\end{minipage} & 2  &  \color{blue}\textbf{2.58} & \color{red}\textbf{3.85}   & 2.82   & 2.82   & 2.76  \\ 

\begin{minipage}{0.1\textwidth}
\centering
\includegraphics[width=0.4\linewidth, height=0.02\textheight]{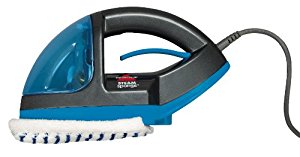}
\end{minipage} &  3 & 2.97  & \color{red}\textbf{4.33}   &2.77   & \color{blue} \textbf{2.99}   &2.96   \\ 

\begin{minipage}{0.1\textwidth}
\centering
\includegraphics[width=0.4\linewidth, height=0.02\textheight]{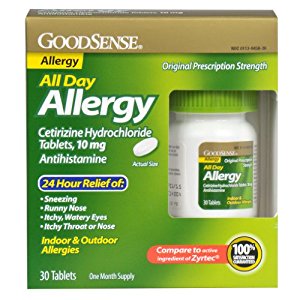}
\end{minipage} &  4 &  4.09  &  \color{red}\textbf{3.68}   & 4.14   & 4.09   &  \color{blue} \textbf{4.04}   \\ 

\begin{minipage}{0.1\textwidth}
\centering
\includegraphics[width=0.4\linewidth, height=0.02\textheight]{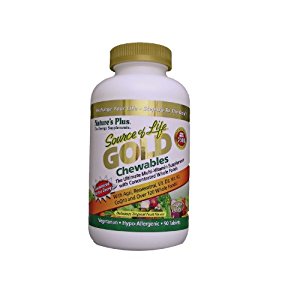}
\end{minipage} &  5 &  \color{blue} \textbf{4.99}  & \color{red}\textbf{4.39}   & 4.47   & 4.94   & \color{blue} \textbf{5.01}   \\ 
 \hline
\end{tabular}
\vskip -0.in
\end{table}

\section{Discussion}
\label{sec:dis}
Empirically, we have shown that multimodal learning (\textbf{+F}) plays an important role in mitigating the problems associated with missing data/modality and, in particular, those associated with the cold-start problem (\textbf{-U} and \textbf{-O}) of recommender systems.  The proposed method LRMM is in line and grounded in recent developments (e.g. DeepCoNN, NRT) to incorporate multimodal data. LRMM distinguishes itself from previous methods: (1) the cold-start problem is reformulated in the context of missing modality; (2) A novel multimodal imputation method which consists of \textit{m}-drop and \textit{m}-auto is proposed to learn models more robust to missing data modalities in both the training and inference stages.   

\section{Related Work}
\label{sec:related_work}
Collaborative filtering (CF) is the most commonly used approach for recommender systems. CF methods generally utilize the item-user feedback matrix. Matrix factorization (MF) is the most popular CF method~\cite{koren2009matrix} due to its simplicity, performance, and high accuracy as demonstrated in previous work~\cite{chen2015matrix}.
Another strength of MF, making it widely used in recommender systems, is that side information other than existing ratings can easily be integrated into the model to further increase its accuracy. Such information includes social network data~\cite{LiWTM15, lagun2015inferring, zhao2016connecting, xiao2017learning}, locations of users and items~\cite{Lu17HBGG} and visual appearance~\cite{he2016ups, mnih2008probabilistic} proposed Probabilistic Matrix Factorization (PMF) which extends MF to a probabilistic linear model with Gaussian noise. 
Following PMF, there are many extensions~\cite{salakhutdinov2007restricted,chen2013general, zheng2016neural, zhang2016discrete, he2016fast, he2017neural} aiming to improve its accuracy.

Unfortunately, CF methods suffer from the cold-start problem when dealing with new items or users without rich information. Content based filtering (CBF)~\cite{pazzani2007content}, on the other hand, is able to alleviate the cold-start problem by taking auxiliary product and user information (texts, images, videos, etc.) into consideration. Recently, several approaches~\cite{almahairi2015learning, xu2014collaborative, he2014predicting, tan2016rating} were proposed to consider the information of review text to address the data sparsity problem which leads to the cold-start problem. The topic model (e.g. LDA~\cite{blei2003latent}) based approaches including CTR~\cite{wang2011collaborative}, HFT~\cite{mcauley2013hidden}, RMR~\cite{Ling2014RMR}, TriRank~\cite{he2015trirank}, and sCVR~\cite{Ren2017} achieve significant improvements compared to previous work on recommender systems. 

Inspired by the recent success of deep learning techniques~\cite{krizhevsky2012imagenet, he2016deep}, some deep network based recommendation approaches have been introduced~\cite{wang2015collaborative, sedhain2015autorec, wang2016collaborative, Seo2017, xue2017deep, zhang2017deep}.  Deep cooperative neural network (Deep-CoNN)~\cite{zheng2017joint} was introduced to learn a joint representation from items and users using two coupled network for rating prediction. It is the first approach to represent users and items in a joint manner with review text. 
TransNets~\cite{catherine2017transnets} extends Deep-CoNN by introducing
an additional latent layer representing the user-item pair. NRT~\cite{Li:2017:NRR} is a method for rating prediction and abstractive tips generation~\cite{dong2017learning}. A four-layer neural network was used for rating regression model. NRT outperforms the state-of-the-art methods on rating prediction.
There is a large body of work for recommender systems and we refer the reader to for surveys of state-of-the-art CF based approaches, CBF methods, and deep learning based methods, respectively~\cite{ShiLH14, LopsGS11, zhang2017deep}.

Our work differs from previous work in that we simultaneously address various types of missing data together with the data-sparsity and cold-start problems. 
\section{Conclusion}
\label{sec:conclusion}

We presented LRMM, a framework that improves the performance and robustness of recommender systems under missing data. LRMM makes novel contributions in two ways: multimodal imputation and jointly alleviating the missing modality, data sparsity, and cold-start problem for recommender systems.  It learns to recommend when entire modalities are missing by leveraging inter- and intra-modal correlations from data through the proposed \textit{m}-drop and \textit{m}-auto methods. LRMM achieves state-of-the-art performance on multiple data sets. Empirically,  we analyzed LRMM in different data sparsity regimes and demonstrated the effectiveness of LRMM. We aim to explore a generalized domain adaptation approach for recommender systems with missing data modalities.

\bibliography{emnlp2018}
\bibliographystyle{acl_natbib}
\end{document}